\documentclass[twocolumn,amsmath,amssymb,prl]{revtex4}
\usepackage{graphicx}
\usepackage{dcolumn}
\usepackage{bm}

\begin{document}

\title{Two-dimensional quantum oscillations of the conductance at LaAlO$_{3}$/SrTiO$_{3}$ interfaces}
\author{A.D. Caviglia$^{1}$, S. Gariglio$^{1}$, C. Cancellieri$^{1}$, B. Sac\'ep\'e$^{1,2}$,  A.F\^ete$^{1}$, N. Reyren$^{1}$, M. Gabay$^{3}$, A.F. Morpurgo$^{1,2}$, J.-M. Triscone$^{1}$}
\affiliation{ $^{1}$D\'epartement de Physique de la Mati\`ere Condens\'ee, University of Geneva, 24 Quai Ernest-Ansermet, 1211 Gen\`eve 4, Switzerland}
\affiliation{ $^{2}$GAP, University of Geneva, 24 Quai Ernest-Ansermet, 1211 Gen\`eve 4, Switzerland}

\affiliation{ $^{3}$Laboratoire de Physique des Solides, Bat 510, Universit\'e Paris-Sud 11, Centre d'Orsay, 91405 Orsay Cedex, France}
\date{\today}

\begin{abstract}
We report on a study of magnetotransport in LaAlO$_{3}$/SrTiO$_{3}$ interfaces characterized by mobilities of the order of several thousands cm$^{2}$/Vs. We observe Shubnikov-de Haas oscillations that indicate a two-dimensional character of the Fermi surface. The frequency of the oscillations signals a multiple sub-bands occupation in the quantum well or a multiple valley configuration. From the temperature dependence of the oscillation amplitude we extract an effective carrier mass $m^{*}\simeq1.45$\,$m_{e}$. An electric field applied in the back-gate geometry increases the mobility, the carrier density and the oscillation frequency.
\end{abstract}

\maketitle

Several experimental and theoretical studies have uncovered a number of remarkable electronic properties of the interface between the complex oxides LaAlO$_{3}$ and SrTiO$_{3}$ \cite{Ohtomo:2004yq,S.Thiel09292006, N.Reyren08312007,Caviglia:2008dq,PhysRevB.81.153414}. One of the main unresolved issues pertains to the dimensionality of the conducting layer \cite{copie:216804}. While it is now clear that, using appropriate growth and annealing conditions, a confined metallic and superconducting electron gas can be formed at such interfaces \cite{Basletic:2008rw,reyren:112506}, no conclusive demonstration of two-dimensional character in the normal state has been obtained so far.

X-Ray absorption spectroscopy experiments \cite{PhysRevLett.102.166804}, as well as density functional theory calculations \cite{PhysRevLett.101.256801}, indicate that an orbital reconstruction occurs at this interface. These studies reveal that, even at room temperature, the degeneracy of the Ti 3$d\text{ }t_{2g}$ levels is lifted and the first available states for the conducting electrons are generated from 3$d_{xy}$ orbitals, which give rise to strongly two-dimensional bands and present a negligible inter-plane coupling. Moreover, the momentum quantization in the quantum well brings about a sub-bands fine structure which was calculated \cite{PhysRevLett.101.256801,PhysRevB.79.245411} but not yet confirmed experimentally. The presence of a very strong spin-orbit coupling adds further complexity to the low-energy electronic structure \cite{PhysRevLett.104.126803,PhysRevLett.104.126802}.

The Fermi surface of two-dimensional electronic states generates clear experimental signatures in the Shubnikov-de Haas (SdH) effect: for a two-dimensional electron gas (2DEG) the quantum oscillations depend only on the perpendicular component of the magnetic field. Previous studies reported quantum oscillations in LaAlO$_{3}$/SrTiO$_{3-\delta}$ heterostructures characterized by a large carrier density (of the order of 10$^{16}$\,cm$^{-2}$) delocalised in the SrTiO$_{3}$ substrate \cite{Ohtomo:2004yq,herranz:216803}. The lack of dependence of the oscillations on the field orientation points to a three-dimensional Fermi surface consistently with previous studies in Nb-doped SrTiO$_{3}$ single crystals \cite{PhysRev.158.775}. Very recently, quantum oscillations with two-dimensional character have been reported for Nb-doped thin films of SrTiO$_{3}$ \cite{Kozuka:2009fq}. However, in LaAlO$_{3}$/SrTiO$_{3}$ heterostructures where the electrons are confined at the interface \cite{copie:216804}, magnetotransport studies have been carried out so far in the diffusive regime, where the scattering time is not sufficiently long to give rise to well defined Landau levels \cite{Brinkman:2007zr, PhysRevLett.104.126803,PhysRevB.80.140403,PhysRevB.80.180410}.

The requirements for observing quantum conductance oscillations are \cite{RevModPhys.54.437}
\begin{equation}
\omega_{\text{c}}\tau>1
\end{equation}
\begin{equation}
\hbar\omega_{\text{c}}>k_{\text{B}}T
\end{equation}
where $\omega_{\text{c}}=e\mu_{0}H/m^{*}$ is the cyclotron frequency, $m^{*}$ is the carrier effective mass, $e$ the elementary charge, $\mu_{0}H$ the applied magnetic field, $T$ is the temperature and $\tau$ is the transport elastic scattering time. To fulfill these conditions with temperatures of the order of 1\,K and magnetic fields of a few Tesla mobilities of $10^{3}$\,cm$^{2}$/Vs or more are required. Existing strategies to reach this value rely on the use of the electrostatic field effect \cite{Caviglia:2008dq,PhysRevLett.103.226802} or defect control \cite{Kozuka:2009fq,Son:2010eq} to improve the quality of the devices. However currently available samples have not yet allowed to access the regime where quantum conductance oscillations are unambiguously visible.

In this paper we present a magnetotransport study performed in LaAlO$_{3}$/SrTiO$_{3}$ interfaces in which the mobility has been boosted by an optimization of the growth conditions, reaching the unprecedented value of 6600\,cm$^{2}$/Vs. In these samples we observe quantum oscillations in the electrical resistance as a function of magnetic field that depend only on the perpendicular component of the magnetic field. An electric field applied to the back-gate modulates the oscillation frequency. These results demonstrate the presence of two-dimensional electronic states originating from quantum confinement at the LaAlO$_{3}$/SrTiO$_{3}$ interface.

\begin{figure}
\begin{center}
\includegraphics[scale=0.23]{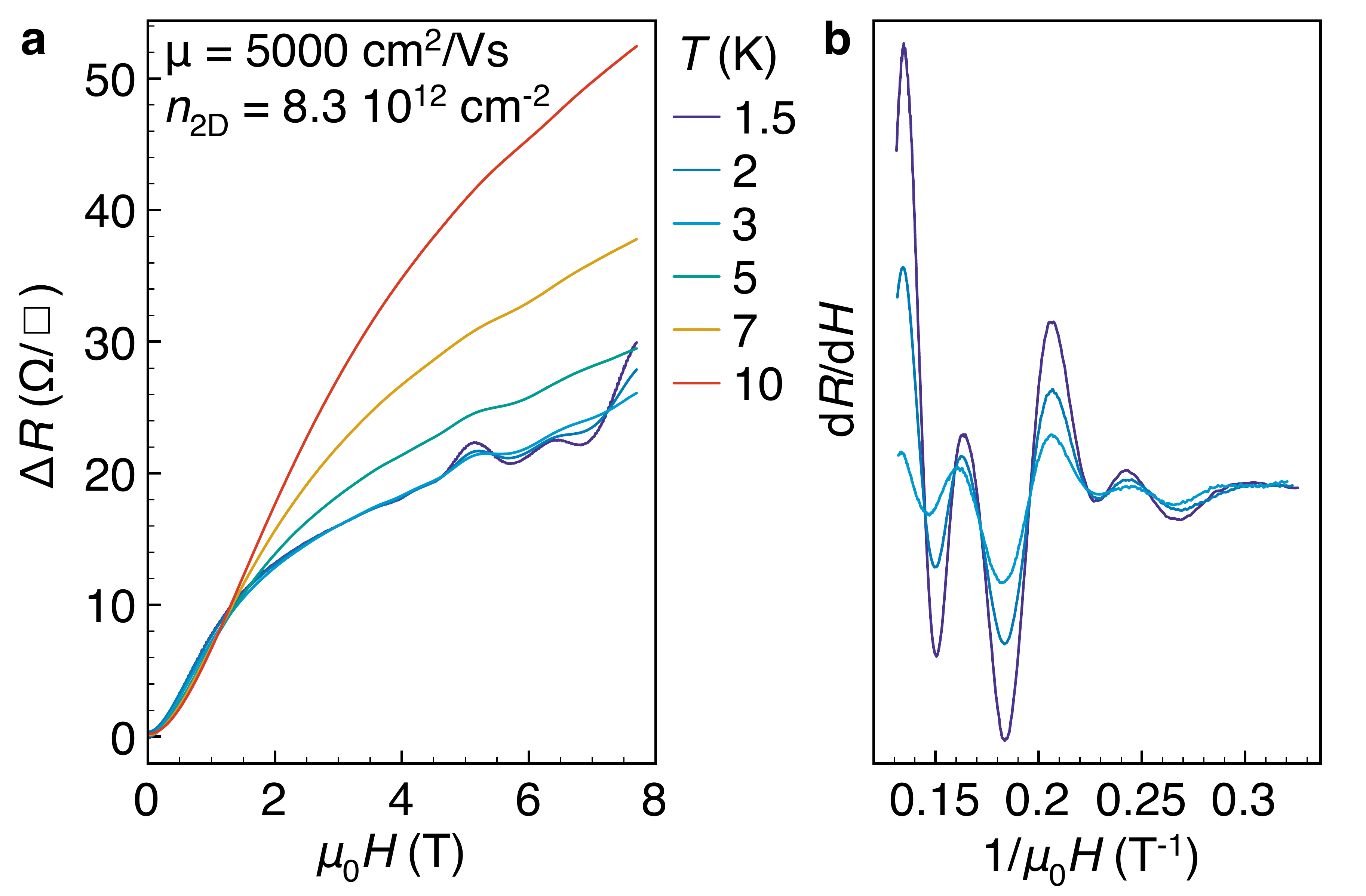}
\caption{\label{fig:oscilla} Shubnikov-de Haas oscillations of the LaAlO$_{3}$/SrTiO$_{3}$ interface. (a) Variation of resistance $\Delta R=R(H)-R(0)$ in response to the application of a magnetic field $H$ oriented perpendicular to the LaAlO$_{3}$/SrTiO$_{3}$ interface, recorded at different temperatures $T$. (b) Numerical derivative d$R$/d$H$ as a function of the inverse of the magnetic field.}
\end{center}
\end{figure}

We fabricated high-mobility samples by depositing epitaxial thin films of LaAlO$_{3}$ on top of TiO$_{2}$-terminated SrTiO$_{3}$ single crystals. The films were grown by pulsed laser deposition at $\sim 650$\,$^{\circ}$C in $\sim 1\times 10^{-4}$ mbar of O$_{2}$ with a repetition rate of 1 Hz. The fluence of the laser pulses was 1 J/cm$^{2}$. The films growth was monitored \textit{in situ} using reflection high energy electron diffraction (RHEED). After growth, each sample was annealed in 200 mbar of  O$_{2}$ at about 530\,$^{\circ}$C for one hour and cooled to room temperature in the same oxygen pressure. The reduced growth temperature, with respect to standard deposition conditions ($\sim 800$\,$^{\circ}$C), significantly improves the crystalline quality of the films, as observed by X-Ray diffraction and RHEED. The samples are metallic and superconducting. They are characterized by mobility values that are approximately one order of magnitude higher than those observed in heterostructures grown with the deposition conditions used in the past.

We have measured a total of 10 heterostructures with a thickness of the LaAlO$_{3}$ top layer between 5 and 10 unit cells, all of which exhibited clear quantum conductance oscillations. Some differences are present in the behavior of the oscillations observed in different samples. In particular the amplitude (and precise shape) of the oscillations varies from sample to sample. In a few samples oscillations are visible directly in the raw magneto-resistance data; in most other samples it is useful to take the derivative of the resistance with respect to the magnetic field --or to remove a background-- to make the oscillations fully apparent. Nevertheless, the overall phenomenology that we discuss here is common to the whole set of devices.

Figure $\ref{fig:oscilla}$a shows the variation of sheet resistance $\Delta R=R(H)-R(0)$ in response to the application of a magnetic field oriented perpendicular to the interface, recorded at different temperatures. The resistance measurements have been performed using a 4 point AC technique, with a current between 10 and 100\,nA, along a transport channel 100\,$\mu$m wide and 200\,$\mu$m long, defined by photolithography. Below $T=7$\,K, oscillations superimposed on a positive background are observed for fields larger than 3\,T, where the regime $\omega_{\text{c}}\tau>1$ is attained (with $\mu=5000$\,cm$^{2}$/Vs, $\omega_{\text{c}}\tau>1$ for $\mu_{0}H>2$\,T). The non-oscillatory background is consistent with previous magnetotransport studies of this system \cite{PhysRevB.80.180410,PhysRevB.80.140403,PhysRevLett.104.126803}. The numerical derivative of the resistance with respect to magnetic field, presented in Figure $\ref{fig:oscilla}$b, reveals that the oscillations are periodic in $1/H$.

\begin{figure}
\begin{center}
\includegraphics[scale=0.23]{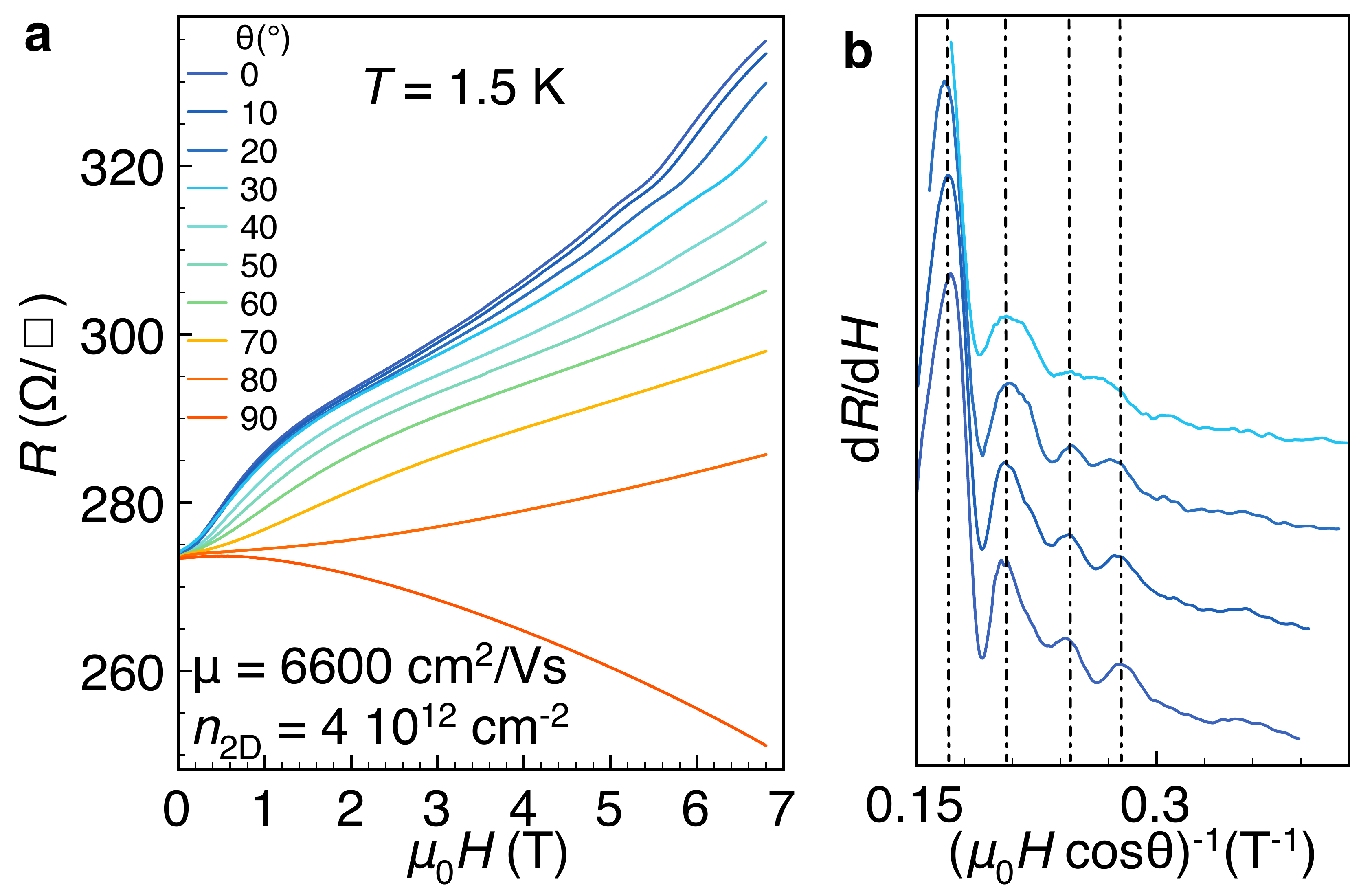}
\caption{\label{fig:oscillangolo}Angular dependence of the quantum oscillations. (a) Sheet resistance $R$ as a function of magnetic field $H$ recorded at different orientations (measured by the angle $\theta$) with respect to the direction normal to the substrate. (b) Numerical derivative d$R$/d$H$ as a function of the inverse of the component of the magnetic field perpendicular to the plane of the interface. An offset has been introduced in each curve for clarity. The lines are a guide to the eye.}
\end{center}
\end{figure}

The dimensionality of the electronic states can be assessed by examining the angular dependence of the quantum oscillations. Figure \ref{fig:oscillangolo}a displays $R(H)$ measured at $T=1.5$\,K on a different sample, for different angles $\theta$. The angle $\theta$ measures the inclination of the magnetic field with respect to the normal to the interface at a fixed azimuthal angle. At $\theta=0^{\circ}$ the magnetic field is applied perpendicular to the interface, while for $\theta=90^{\circ}$ the magnetic field vector lies in the plane of the interface, parallel to the current. At $\theta=90^{\circ}$ we observe, consistently with previous reports, a fairly large negative magnetoresistance \cite{PhysRevB.80.140403,PhysRevLett.104.126803}. Figure \ref{fig:oscillangolo}b shows the derivative of the data from 0$^{\circ}$ to 30$^{\circ}$ as a function of $(H\cos\theta)^{-1}$. It is apparent that the oscillation depends only on the perpendicular component of the magnetic field. This observation directly indicates that the oscillations arise from closed orbits in momentum space along a two-dimensional Fermi surface.

\begin{figure}
\begin{center}
\includegraphics[scale=0.23]{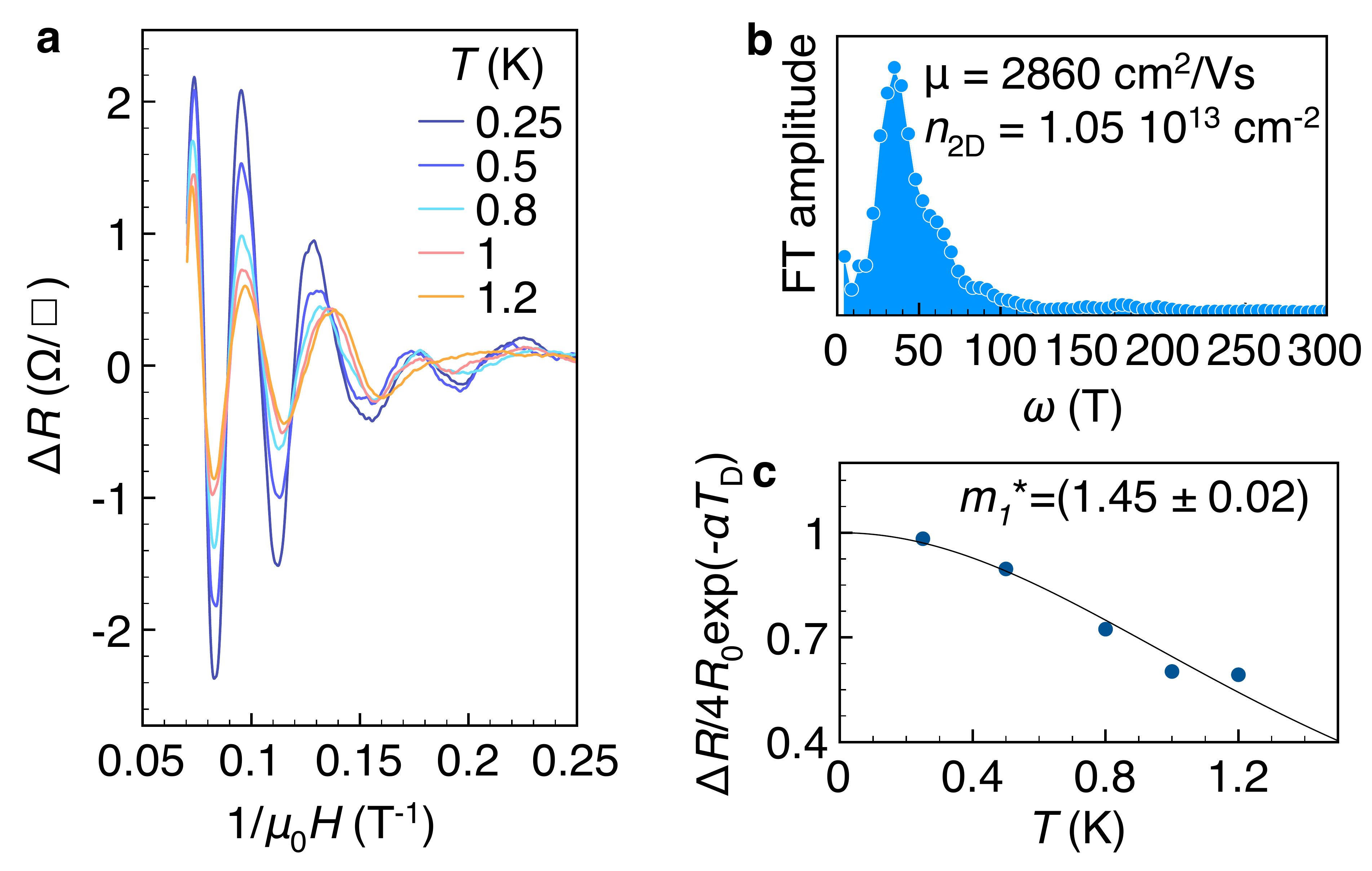}
\caption{\label{fig:oscilltemp}Temperature dependence of the Shubnikov-de Haas oscillations. (a) Oscillatory component of the sheet resistance $\Delta R$ as a function of the inverse of the magnetic field oriented perpendicular to the LaAlO$_{3}$/SrTiO$_{3}$ interface, recorded at different temperatures $T$. (b) Fourier spectrum of the resistance oscillation measured at 250\,mK. (c) Blue dots: temperature dependence of the amplitude of the oscillation at 13.5\,T. Solid line: best fit to equation \ref{eq:th} of the experimental data}
\end{center}
\end{figure}

SdH oscillations represent a direct measurement of the area of the Fermi surface. Indeed, the Onsager relation $\omega=(\phi_{0}/2\pi^{2})A$ establishes a direct proportionality between the cross-sectional area of the Fermi surface normal to the magnetic field, $A$, and the frequency of the oscillation $\omega$ ($\phi_{0}$ is the magnetic flux quantum) \cite{ashcroft}. Figure  \ref{fig:oscilltemp}b displays the Fourier transform of the oscillatory component of the magnetoresistance of a sample characterized by a mobility of the order of 3000\,cm$^{2}$/Vs and a Hall carrier density $n_{\text{2D}}=1.05\cdot 10^{13}$\,cm$^{-2}$. A main broad peak is clearly visible in the Fourier transform at $35$\,T with a shoulder at 50\,T. In other samples the shoulder appears as a small secondary peak. The main peak leads to an area of $0.33$\,nm$^{-2}$, which is only 0.1\% of the Brillouin zone. Assuming a circular section of the Fermi surface, we can estimate the carrier density as $n_{\text{2D}}=\omega\nu_{v}\nu_{s}e/h$, where $\nu_{v}$ and $\nu_{s}$ indicate the valley and spin degeneracy respectively. By taking  $\nu_{s}=2$ and a single valley we find $n_{\text{2D}}=1.69\cdot 10^{12}$\,cm$^{-2}$ (the feature at 50\,T would give an additional contribution of $2.4\cdot 10^{12}$\,cm$^{-2}$). The corresponding Fermi wavelength is $\lambda_{\text{F}}=2\pi/k_{\text{F}}=19$\,nm. This value is larger than the thickness of the electron gas at low temperatures ($\simeq 10$\,nm) estimated in low-mobility samples using the anisotropy of the superconducting critical fields \cite{reyren:112506}, infrared ellipsometry \cite{PhysRevLett.104.156807} and atomic force microscopy \cite{Basletic:2008rw, copie:216804}. Apart from the apparent discrepancy between the density values extracted from the Hall effect measurements and the SdH oscillations (which will be discussed later), these findings are consistent with a picture of a two-dimensional electron gas formed by quantum confinement at the LaAlO$_{3}$/SrTiO$_{3}$ interface.

We now turn our attention to the temperature dependence of the oscillations amplitude, which contains important information regarding the carrier effective mass and the level of disorder. For this analysis we performed magnetoresistance measurements up to 15\,T and down to 250\,mK. The amplitude of the quantum oscillation, extracted from these measurements by subtracting a polynomial background (see Figure \ref{fig:oscilltemp}a), decreases with increasing temperature. 

The oscillations amplitude $\Delta R$ as a function of temperature $T$ can be described by the relation \cite{RevModPhys.54.437}
\begin{equation}\label{eq:th}
\Delta R (T)=4R_{0} e^{-\alpha T_{\text{D}}}\alpha T/\sinh(\alpha T)
\end{equation}
where $\alpha=2\pi^{2}k_{\text{B}}/\hbar\omega_{\text{c}}$, $R_{0}$ is the non-oscillatory component of the square resistance and $T_{\text{D}}$ is the Dingle temperature. The best fit to equation \ref{eq:th} of our experimental data, is shown in Figure \ref{fig:oscilltemp}c, for the largest maximum of the oscillations (the second largest gives the same result). The fitting parameters are the carrier effective mass (which enters the cyclotron frequency) and the Dingle temperature. We observe a good agreement between theory and experiment with $m^{*}=1.45\pm 0.02$ $m_{e}$ and $T_{D}=6$\,K. Our estimation of the effective mass is similar to what has been observed in doped SrTiO$_{3}$ by quantum oscillations in thin films \cite{Kozuka:2009fq} and by optical conductivity in crystals \cite{mechelen:226403}. We note also that in samples with an order of magnitude larger carrier density, infrared ellipsometry gives an effective mass that is a factor of 2 larger than what we find \cite{PhysRevLett.104.156807}. Finally, the broadening of the Landau levels that we find, determined by $k_{\text{B}}T_{D}\sim 0.5$\,meV, is not much smaller than the their energy separation ($\hbar\omega_{\text{c}}\sim 1$\,meV at 10\,T). This explains the small amplitude of the oscillations that we observe.

\begin{figure}
\begin{center}
\includegraphics[scale=0.23]{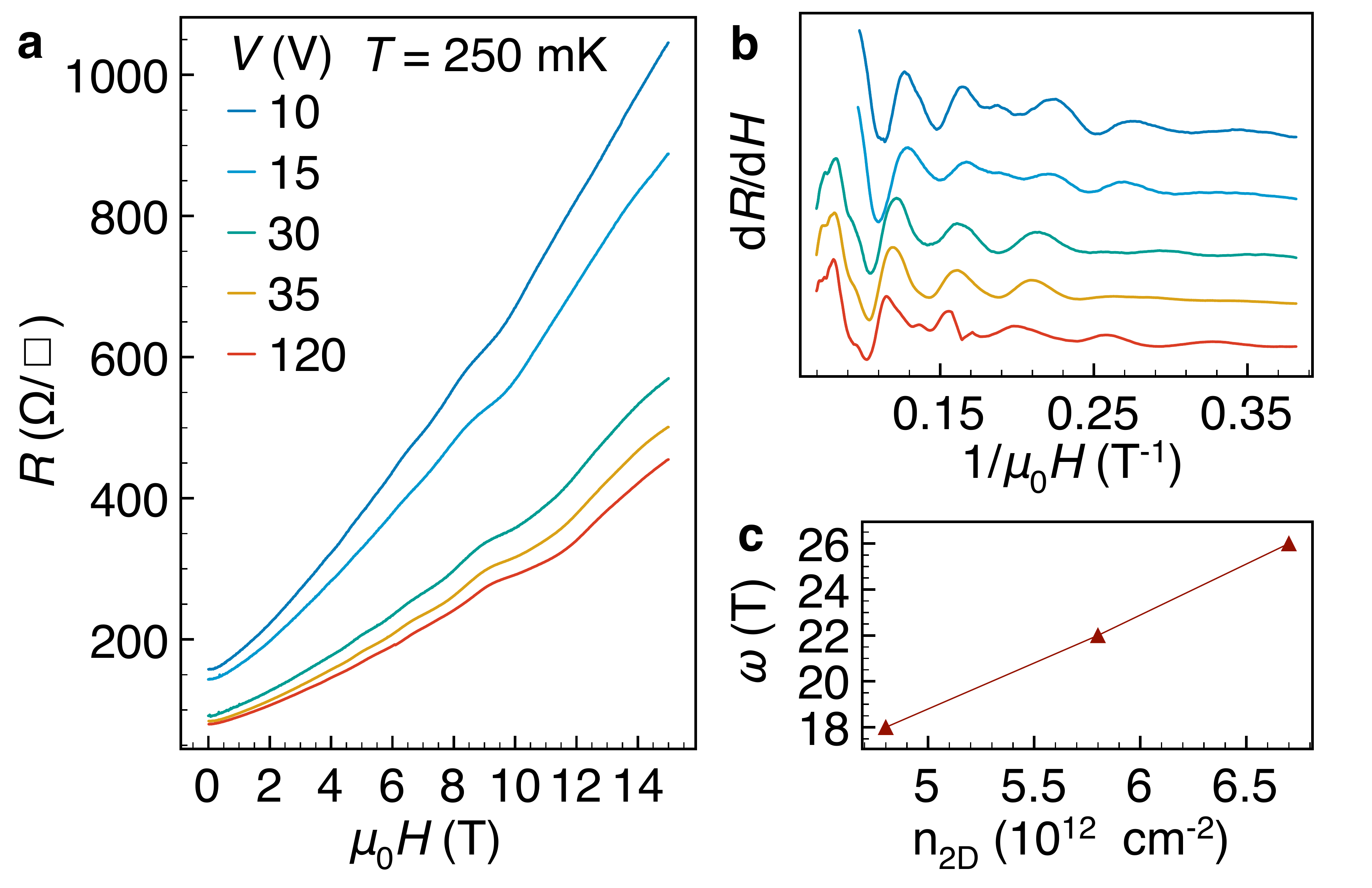}
\caption{\label{fig:vg} Field effect modulation of the Shubnikov-de Haas oscillations. (a) Sheet resistance $R$ as a function of magnetic field $H$ recorded at different gate voltages $V$ at $T=250$\,mK. (b) Numerical derivative d$R$/d$H$ as a function of the inverse of the magnetic field showing the effect of the gate voltage on the resistance oscillations. (c) Main frequency of the oscillation $\omega$ as a function of carrier density $n_{\text{2D}}$.}
\end{center}
\end{figure}

On some of our samples we have also looked at the evolution of the quantum oscillations in devices equipped with a gate electrode, which consists of a metal layer deposited on the back side of the substrate (0.5\,mm thick). Figure \ref{fig:vg}a,b displays $R(H)$ measurements and d$R$/d$H(H^{-1})$ recorded at $T=250$\,mK and different gate voltages $V$. In conventional two-dimensional electron gases, the main effect of tuning the gate voltage is to modulate the carrier density, leading to a change in the Fermi surface, and thereby in the SdH oscillation frequency.  Indeed, as expected, we observe a clear shift of the main peak of the oscillations (see Fig. \ref{fig:vg}c), whose frequency increases linearly with carrier density \footnote{Interestingly, the smaller secondary peak at $\simeq 50$\,T does not seem to shift significantly with gate voltage}. In performing these experiments, we also noted that in most devices the oscillations become more apparent at larger gate voltages. This is a consequence of the fact that the mobility increases with $V$ \cite{PhysRevLett.103.226802} (in the device whose data are shown in Fig. \ref{fig:vg} the mobility increases from 2400\,cm$^{2}$/Vs at 10\,V to 3600\,cm$^{2}$/Vs at 120\,V).

We conclude that the realization of high-quality LaAlO$_{3}$/SrTiO$_{3}$ interfaces enables us to reveal the presence of discrete two-dimensional sub-bands that manifests themselves in SdH conductance oscillations. One issue that remains to be solved is the apparent mismatch between the values of the carrier density estimated from the Hall effect and from the SdH oscillations (approximately a factor of 4-5 if we consider only the main peak, a factor of 2 if we consider also the peak at 50\,T). Apart from the possibility that the limited number of periods observed experimentally may not be sufficient to resolve all peaks in the spectrum of the SdH oscillations, several physical scenarios are also possible. First, even though band-structure calculations predict the presence of only one valley \cite{PhysRevLett.101.256801} in the conduction band of LaAlO$_{3}$/SrTiO$_{3}$ interfaces, it cannot be excluded that in reality multiple valleys are present. Second, effects associated to the spin of the electrons may be important. On the one hand, a strong spin-orbit interaction is known to affect SdH oscillations \cite{PhysRevLett.104.126803,PhysRevLett.78.1335}. On the other hand, since the effective mass is larger than the free electron mass, the Zeeman splitting is larger than the Landau level spacing even when the gyromagnetic factor is $g=2$, which brings the system in an unconventional regime. Finally, the different sub-bands occupied by electrons in the 2DEG may originate from different $d$ orbitals, leading to different effective masses, mobility values, and scattering times (which would not fulfill the condition to observe the oscillations) \cite{PhysRevB.79.245411}. Additional studies are required to elucidate this point, which most likely will require samples of even higher quality and higher magnetic fields.

The observation of two dimensional sub-bands in LaAlO$_{3}$/SrTiO$_{3}$ interfaces establishes a connection between this system and more conventional semiconducting heterostructures based on III-V compounds. In contrast to these more conventional systems 2DEGs at LaAlO$_{3}$/SrTiO$_{3}$ interfaces are characterized by very strong spin-orbit interaction, much lower Fermi energy, higher effective mass, not to mention the occurrence of superconductivity. The availability of high-quality LaAlO$_{3}$/SrTiO$_{3}$ interfaces give access to two-dimensional electron physics in an entirely unexplored parameter regime, which is likely to disclose new phenomena not yet observed in the diffusive transport limit that has been investigated so far.

We thank T. Giamarchi and D. van der Marel for useful discussions and Marco Lopes for his technical assistance. We acknowledge financial support by the SNSF through the NCCR ``Materials with Novel Electronic Properties" MaNEP and Division II, by the EU through the projects ``Nanoxide" and ``OxIDes".

\end{document}